\documentstyle[aaspp4,flushrt,psfig]{article}

\newcommand{\msun}{M_\odot}

\lefthead{N. Wex, V. Kalogera, \& M. Kramer}
\righthead{Constraints on Supernova Kicks}

\begin{document}

\title{Constraints on Supernova Kicks from the Double Neutron Star System PSR 
B1913+16}

\author{N. Wex \altaffilmark{1,2}, V. Kalogera \altaffilmark{3}, and
M. Kramer \altaffilmark{4,1}}

\altaffiltext{1}{MPI f\"ur Radioastronomie, Auf dem H\"ugel 69,
                 53121 Bonn, Germany}
\altaffiltext{2}{Joseph Henry Laboratories and Physics Department,
                 Princeton University, Princeton, New Jersey 08544, USA}
\altaffiltext{3}{Harvard-Smithsonian Center for Astrophysics, 60 Garden
St., Cambridge, MA 02138, USA; vkalogera@cfa.harvard.edu}
\altaffiltext{4}{Astronomy Department, 601 Campbell Hall, University of
                 California, Berkeley, California 94720, USA}

\begin{abstract}

We use recent information on geodetic precession of the binary pulsar B1913+16
along with measurements of its orbital parameters and proper motion to derive
new constraints on the immediate progenitor of this double neutron star
system. As part of our analysis we model the motion of the binary in the
Galaxy after the second supernova explosion, and we derive constraints on the
unknown radial velocity.  We also obtain limits on the magnitude and direction
of the kick velocity imparted to the pulsar companion during the second
supernova explosion. We consider the complete set of possible cases, depending
on the kinematic age of the system and the 180$^\circ$ ambiguity in the pulsar
spin orientation.  Most interestingly, we find that the natal kick must have
been directed almost perpendicular to the spin axis of the neutron star
progenitor, independent of the specific pre-supernova configuration. Such a
tight constraint on the kick direction has important implications for the
physical mechanism responsible for the kick.

\end{abstract}

\section{Introduction}

Stellar systems with one or more compact objects can provide us with valuable
information regarding the collapse of massive ($\gtrsim 8-10$ M$_\odot$) stars
and the formation of neutron stars (NSs) and black holes.  Basic issues such
as the relation of progenitor mass to the type and mass of the compact object
as well as the effects of mass ejection or fallback are still poorly
understood. Furthermore, despite the accumulation of pieces of evidence for
kicks imparted to NSs at birth in recent years (see \cite{vdHvP97}1997;
\cite{Bur98}1998; and references therein) the physical origin of kicks and the
factors determining its magnitude and direction remain unknown.

Among the different types of binaries with compact objects, double NS systems
offer important advantages in studying NS progenitors and kicks, mainly
because after the second supernova (SN) explosion these systems have not
evolved in any way other than through energy and angular momentum loss due to
gravitational wave emission. Therefore their present observed properties can
be related to the conditions just prior to the second SN in a straightforward
way and can be used to constrain the parameters of the pre-SN system and the
kick imparted to the second NS.  Such constraints on the progenitors of double
NS systems have the capability of testing formation and evolutionary sequences
for binary systems and thus provide input to models for the population of
Galactic double NS binaries. This is of particular interest, since merging
double NS systems are currently considered as the most promising sources of
gravitational waves (\cite{Schut86}1986) and a possible origin of $\gamma$-ray
bursts (\cite{Pac86}1986).

The double NS system containing the radio pulsar PSR B1913+16, which was
discovered in 1974 by Hulse and Taylor (\cite{HT75}1975), proved to be the
most exciting test laboratory for relativistic theories of gravity.  General
relativity passed this test with flying colors and provided us with detailed
information about the system. Continuous timing observations of PSR B1913+16
over a period of twenty years (\cite{Tay&al76}1976; \cite{TFM79}1979;
\cite{TW82}1982; \cite{TW89}1989; \cite{Tay94}1994) allowed a highly accurate
determination of the spin parameters, the pulsar position on the sky, the five
Keplerian parameters of the orbital motion, and three post-Keplerian
parameters related to effects predicted by general relativity. The measured
values for the three post-Keplerian parameters are in excellent agreement with
general relativity and allow a precise determination of the (observed) masses
of the pulsar and its companion, $M_p$ and $M_c$, the semi-major axis of the
(relative) orbit, $A$, and the inclination of the orbital plane with respect
to the plane of the sky, $i$. The proper motion of the pulsar on the sky has
also been measured with moderate accuracy (\cite{DT91}1991). In addition to
spin, astrometric and orbital parameters, timing observations at different
frequencies has given a precise value for the dispersion measure, DM, of the
pulsar, which can be converted into a distance given a model for the
distribution of free electrons along the line-of-sight. Table \ref{P1913}
lists all the measured and derived parameters for the PSR B1913+16 binary
system relevant to the present study.

Reports about changes in the relative amplitudes of the prominent leading and
trailing components of the integrated profile of the pulsar by
\cite{WRT89}(1989) and \cite{CWB90}(1990) have been explained as a result of a
change in the direction of the pulsar spin caused by geodetic precession.
This requires a misalignment between the orbital angular momentum and the spin
of the pulsar. Based on recent observations with the 100-m Effelsberg radio
telescope, \cite{Kra98}(1998) also detected a decrease in the relative
separation of these components, strengthening the case for geodetic
precession.  Assuming a circular hollow-cone like beam, \cite{Kra98}(1998) was
able to derive a value for the misalignment angle of, i.e.~$\Theta =
22(-4/+2)^\circ$ or $\Theta = 158(-2/+4)^\circ$. More recently
\cite{Tay99}(1999) used pulse structure and polarization data measured with
the 305m Arecibo radiotelescope and obtained $\Theta=14(\pm 2)^\circ$ or
$\Theta=166(\pm 2)^\circ$ as the most likely inclination of the pulsar spin
axis with a safe upper (lower) limit of $22^\circ$ ($158^\circ$).  In what
follows we adopt this upper (lower) limit, since the limits on the pre-SN and
kick parameters derived here become tighter for smaller (larger) values of
$\Theta$.

Given our general understanding of the evolutionary history of double NS
systems, the spins of the binary components and the orbital angular momentum
axis before the second SN were aligned due to mass transfer in the binary. The
inferred misalignment then between the pulsar spin and orbital angular
momentum has been caused by the supernova kick given to the second NS and
corresponds to the angle between the pre-SN and post-SN orbital planes.  With
this new piece of information about the misalignment angle and the proper
motion measurements at hand we extend the work by \cite{FK97}(1997) and
combine all available data for the PSR B1913+16 binary system. We analyze the
motion of the system in the Galaxy and the effect of asymmetric SN explosions
on binary orbits and we are able to constrain the properties of its immediate
progenitor as well as its Galactic velocity and the actual birth kick
(magnitude and direction) imparted to the pulsar companion during the second
SN explosion. The constraints on the kick direction have important
implications for the physical mechanism responsible for the SN kick.

In the next section we discuss the position and motion of PSR B1913+16 in the
Galactic gravitational potential in order to estimate its kinematic age and
place of birth. In Section \ref{Relations} we give relations between the
present binary system and its immediate progenitor. These relations are
applied in the final section followed by a discussion.


\section{Motion in the Galaxy \label{Motion}}

The velocity of the center-of-mass of a binary system after a SN explosion is
determined by properties of the kick imparted to the newly-born NS and the
pre-SN binary parameters.\footnote[7]{Referring to pre- and post-SN systems in
the following, we always mean the system immediately before and just after the
{\em second} SN explosion.} Here, we estimate the center-of-mass velocity,
$V_{\rm CM}$, of PSR B1913+16 after the second SN explosion by making use of
its measured proper motion. The proper motion of the system at present is
related to a number of different velocity components in addition to the NS
kick velocity. The post-SN velocity is the vector sum of the circular speed
due to Galactic rotation at the SN location, of the peculiar velocity with
respect to this local velocity, and of the velocity imparted to the system due
to the NS kick and the mass loss associated with the explosion. Depending on
the age of the system this velocity sum is further altered by the
gravitational acceleration in the Galactic potential, so its velocity relative
to the present Galactic environment may not be representative of the velocity
of the system right after the second SN explosion.

Given that the details of the complete evolutionary history of PSR B1913+16
are unknown, to make progress in estimating the post-SN $V_{{\rm CM}}$, we
make a few simplifying assumptions, which, however, are consistent with our
qualitative understanding of the formation of double NS systems
(\cite{SvdH82}1982; \cite{BvdH91}1991; \cite{LPP96}1996; \cite{Bag97}1997;
\cite{PZY98}1998).  The initial progenitor of PSR B1913+16 is thought to be a
binary with two massive stars. At the time of the first SN explosion the
system was a relatively wide binary with a typical orbital velocity of a few
10 km s$^{-1}$. The corresponding center-of-mass velocity imparted to the
system at the first explosion is of a similar magnitude
(\cite{Kal96}1996). Therefore, we assume that the progenitor was in the
Galactic plane at the time of the second SN explosion and that its peculiar
velocity was small compared to the local Galactic rotational velocity ($\sim
200$ km s$^{-1}$; see also \cite{BP95}1995).  The observed pulsar is the older
of the two NSs, created in the first SN explosion, and has been recycled
during an accretion phase.  We further consider the characteristic age of the
pulsar ($\tau_c\equiv P/2\dot{P}\simeq 109$\,Myr) to be an upper limit on the
time, $T_{\rm SN}$, elapsed since the second SN explosion.  In fact,
\cite{ACW98}(1998) derived an upper limit of 80 Myr by estimating the
equilibrium period of the pulsar during the accretion phase (spin-up limit
$\tau_{{\rm su}}$).

We take as a model for the Galactic potential the model of \cite{KG89}(1989)
(see Appendix A) and integrate the position and velocity of PSR B1913+16 back
in time from the present values to the time of the second SN explosion.  We
introduce a Cartesian coordinate system $(X,Y,Z)$ with the Galactic center at
its origin, the $Z$-axis perpendicular to the Galactic plane, and the Sun at
$(X_\odot,0,Z_\odot)$, where $X_\odot=-8.5\pm0.5$ kpc and $Z_\odot\simeq 15$
pc (\cite{FW97}1997; \cite{Coh95}1995). The $Y$-axis of our coordinate system
is at a Galactic longitude $l=90^\circ$ and in direction of the Galactic
rotation.

The determination of the pulsar's present location in the Galaxy requires its
position on the sky and its distance from the solar system.  While the
Galactic longitude, $l$, and the Galactic latitude, $b$, are known with good
precision, $l=49.968^\circ$ and $b=2.122^\circ$, the only distance indicator
available at present is the dispersion measure based on a model for the
Galactic distribution of free electrons. \cite{DT91}(1991) give $d=8.3\pm1.4$
kpc as a distance to PSR B1913+16. The model of \cite{TC93}(1993) gives a
distance of 7.1 kpc which lies within the error bars given above. For a
distance of 8.3 kpc the present position of PSR B1913+16 in the Galaxy, ${\bf
X}_P$, is $(-3.16,6.35,0.33)$ in units of kpc. Hence, at present the pulsar is
about 300 pc above the Galactic plane.

The proper motion of PSR B1913+16 in right ascension and declination is
known with about 10\%
accuracy (\cite{DT91}1991):
\begin{eqnarray}
&& \mu_\alpha=\dot\alpha\cos\delta=-3.27\pm0.35\;{\rm mas}\;{\rm yr}^{-1}\;,
\\
&& \mu_\delta=\dot\delta          =+1.04\pm0.42\;{\rm mas}\;{\rm yr}^{-1}\;.
\end{eqnarray}
The corresponding proper motion in Galactic longitude and latitude is 
\footnote[6]{$\mu_\alpha$ and $\mu_\delta$ have a normalized covariance of
+0.3147.}
\begin{eqnarray}
&& \mu_l = \dot l\cos b = -0.60\pm0.45\;{\rm mas}\;{\rm yr}^{-1}\;, \\
&& \mu_b = \dot b       = +3.38\pm0.31\;{\rm mas}\;{\rm yr}^{-1}\;.
\end{eqnarray}
The radial velocity of the binary, $V_r$, with respect to the solar system
cannot be measured or inferred from pulsar-timing observations.  Consequently,
we treat it here as a free parameter ($V_r>0$ means that the pulsar is moving
away from us), although our calculations allow us to set limits on its
magnitude.

For a specific assumed value of the radial velocity, we obtain the present
motion of PSR B1913+16 in an inertial system co-moving with the solar
system. Its velocity in the Galactic ($X$-$Y$-$Z$) frame is then simply
\begin{equation}
   {\bf V}_P = {\bf V}_\odot + V_r
   \left(\begin{array}{c}
      \cos b \cos l \\
      \cos b \sin l \\
      \sin b
   \end{array}\right)
   +\mu_l d
   \left(\begin{array}{c}
      -\sin l \\
      \cos l \\
           0 
   \end{array}\right)
   +\mu_b d
   \left(\begin{array}{c}
      -\sin b \cos l \\
      -\sin b \sin l \\
       \cos b
   \end{array}\right) \;.
\end{equation}
The Galactic velocity of the Sun is the Galactic velocity of the local
standard of rest (see Appendix A) plus the peculiar velocity of the Sun
(\cite{DB97}1997; \cite{Bie99}1999):
\begin{equation}
   {\bf V}_\odot = \left(
   \begin{array}{r}
   {\rm   0 \; km/s} \\
   {\rm 221 \; km/s} \\
   {\rm   0 \; km/s}
   \end{array}\right)
   +\left(
   \begin{array}{r}
   {\rm  10 \; km/s} \\
   {\rm   5 \; km/s} \\
   {\rm   7 \; km/s}
   \end{array}\right) \;.
\end{equation}

Given the present position and velocity, ${\bf X}_p$ and ${\bf V}_p$, we
calculate the pulsar's Galactic trajectory back in time using a model for the
gravitational potential of the Galaxy (Appendix A). For $V_r \lesssim -500$
km/s or $V_r \gtrsim 200$ km/s, we find {\em only} one intersection with the
Galactic plane ($Z=0$ pc) just a few Myrs in the past (case A). For smaller
radial velocities one finds several additional intersections, a second one
$\gtrsim 35$ Myrs in the past (case B) and even a third one $\gtrsim 135$ Myrs
in the past (case C). These kinematic ages, $\tau_{{\rm kin}}$, as a function
of $V_r$ are shown in Fig.~\ref{figage}. The third intersection is well beyond
the age inferred from the spin-up limit, $\tau_{{\rm su}}$, and the
characteristic age of the pulsar, $\tau_{{\rm c}}$, and therefore we exclude
this case from our analysis.

Given our assumption that the progenitor was in the Galactic plane at the time
of the second SN, we take the position and velocity of the intersections to
correspond to the post-SN conditions.  In what follows we consider two
separate cases.  In the first case (A), the binary pulsar has intersected the
Galactic plane only once, i.e.~we are seeing the direct runaway of the binary
pulsar out of the Galactic plane, caused primarily by the kick imparted to the
system at the last SN. In this case its kinematic age is $\sim3$ Myr for a
wide range of radial velocities (see Fig.~\ref{figage}).  In the second case
(B), the binary pulsar has moved from below to above the Galactic plane.  Note
that in this case, given the oscillatory nature of the motion perpendicular to
the Galactic plane, the time spent by the binary system close to the Galactic
plane ($|z|\la300$ pc) at its present position, is short compared to its
kinematic age ($<8\%$; see Fig.~\ref{figz}).  Consequently, we consider case
A as more likely.  Nevertheless, we will investigate both cases.

If the binary system is indeed old, i.e.\ case B, then there is a lower limit
to its age of $\sim35$ Myr (c.f.\ Fig.~\ref{figage}).  The minimum age of 60
Myr derived by \cite{ACW98}(1998) is a result of their assumption of constant
Galactocentric radius for the motion of the B1913+16 system, which is expected
to brake down for these long ages.

We obtain the magnitude and direction of the center-of-mass velocity, ${\bf
V}_{{\rm CM}}$, imparted to the system due to the second SN explosion by
subtracting the Galactic rotation velocity from the velocity of the binary
system at $Z=0$ pc (under the assumption that any pre-SN peculiar velocity
relative to Galactic rotation was small). The magnitude of ${\bf V}_{{\rm
CM}}$ is given in Fig.~\ref{figvsys} as a function of the radial velocity
$V_r$., for both cases A and B. In what follows we use this magnitude as a
constraint on the properties of the SN kick and of the pre-SN binary system.

It is worth noting that none of the calculated pulsar trajectories in
the Galaxy penetrates far into the central region, where the axisymmetric
model for the gravitational potential might fail to be a good approximation.


\section{Orbital Dynamics at the Supernova Explosion \label{Relations}}

In double NS systems the compactness of the two binary members excludes any
mass transfer or tidal interaction between them after the second SN. Any
evolution of the binary characteristics is due only to the emission of
gravitational waves. For a given radial velocity, $V_r$, we calculate the
time, $T_{{\rm SN}}\equiv\tau_{{\rm kin}}$, elapsed since the second SN, i.e.\
the kinematic age of the binary system (see Fig.~\ref{figage}) and use the
(first-order) equations of \cite{Pet64}(1964) to calculate the post-SN
semi-major axis, $A_f$, and eccentricity, $e_f$.  As discussed in more detail
in \cite{FK97}(1997), the various evolutionary sequences discussed in the
literature for the formation of double NS converge to the same configuration
before the second SN: a binary consisting of a NS and a helium star in a
circular orbit (because of an earlier common-envelope or mass transfer phase
that led to the recycling of the pulsar). We analyze the effect of the SN
explosion to the orbital dynamics of the system and we relate the post-SN
parameters (masses, orbital size and eccentricity, center-of-mass velocity,
misalignment angle between spin and orbital angular momentum) to the pre-SN
parameters (helium-star mass and orbital size) as well as to the magnitude and
direction of the SN kick. Effects of the expanding SN shell on the compact
companion are absolutely negligible (c.f.\ calculations in \cite{FA81}(1981)
and \cite{YSN93}(1993), where the effective cross section of the NS companion
is given by its radius of Bondi-Hoyle accretion).

The (circular) pre-SN PSR B1913+16 binary system is characterized by the mass
of the exploding star, $M_i$, the mass of the visible pulsar, $M_p$, and their
separation, $A_i$. The relative orbital velocity, $V_0$, between $M_i$ and
$M_p$
\begin{equation}
   V_0 = \sqrt{\frac{G(M_i+{M_p})}{A_i}} \;,
\end{equation}
where $G$ denotes Newton's gravitational constant.  The post-SN binary system
is characterized by the mass of the NS born in the SN explosion, $M_f\equiv
M_c$, the mass of the recycled pulsar, $M_p$, the semi-major axis of the
binary orbit, $A_f$, its orbital eccentricity, $e_f$, and the angle $\theta$
between the orbital angular momenta of pre- and post-SN orbits. The parameters
of the pre- and post-SN binary systems determine the kick velocity ${\bf V}_K$
imparted to the NS at the moment of its birth in the SN explosion.  Using
equations in \cite{Hil83}(1983) and \cite{Kal96}(1996) we find
\begin{equation}
\left.\begin{array}{lcl}\label{vxvyvz}
   v_x^K &=& \displaystyle\pm\eta
             \left[\left(1-\frac{1}{\alpha(1+e_f)}\right)
                   \left(\frac{1}{\alpha(1-e_f)}-1\right)\right]^{1/2} \\[3mm]
   v_y^K &=& \eta\cos\theta-1 \\[3mm]
   v_z^K &=& \eta\sin\theta 
\end{array}\right\} 
\quad \eta\equiv\sqrt{\alpha\beta(1-e_f^2)} \;,
\end{equation}
where we used following definitions:
\begin{equation}
   \alpha    \equiv \frac{A_f}{A_i}\;,\qquad
   \beta     \equiv \frac{M_c+M_p}{M_i+M_p}\;,\qquad
   {\bf v}_K \equiv \frac{{\bf V}_K}{V_0}\;.
\end{equation}

During the SN explosion, the post-SN system as a whole receives a velocity
relative to the center of mass of the pre-SN binary. The magnitude of this
center-of-mass velocity ${\bf V}_{{\rm CM}}$ (${\bf v}_{{\rm CM}} \equiv {\bf
V}_{{\rm CM}}/V_0$) is given in \cite{Kal96}(1996):
\begin{equation}\label{vsvs}
   v_{{\rm CM}}^2 = \kappa_1+\kappa_2\left(2-\frac{1}{\alpha}\right)
                     -\kappa_3\sqrt{\alpha(1-e_f^2)}\;\cos\theta \;,
\end{equation}
where 
\begin{equation}
   \kappa_1\equiv\frac{M_i^2}{(M_i+M_p)^2} \;,\qquad 
   \kappa_2\equiv\frac{M_c^2}{(M_c+M_p)(M_i+M_p)} \;,\qquad 
   \kappa_3\equiv2\sqrt{\kappa_1\kappa_2} \;.
\end{equation}
If ${\bf k}$ denotes the the unit vector in direction of the orbital
angular momentum of the post-SN system, then
\begin{equation}\label{vsp}
   v_{{\rm CM}}^\| \equiv {\bf v}_{{\rm CM}} \cdot {\bf k} =
   \sqrt{\kappa_1}\;\sin\theta \;.
\end{equation}
For the angle between ${\bf v_{{\rm CM}}}$ and the orbital angular momentum of
the post-SN orbit we find
\begin{equation}
   \alpha_v = \arccos(v_{{\rm CM}}^\|/v_{{\rm CM}}).
\end{equation}
The calculations described in section 2 provide the angle between the 
line-of-sight, ${\bf K}_0$, and ${\bf v_{{\rm CM}}}$:
\begin{equation}
   i_v = \arccos({\bf K}_0\cdot{\bf v_{{\rm CM}}}).
\end{equation}
The angles $\alpha_v$, $i_v$, and $i$ form a spherical triangle and thus
satisfy the condition
\begin{equation}\label{cond1}
   -1\le\frac{\cos i-\cos i_v\cos\alpha_v}{\sin i_v\sin\alpha_v}\le1 \;.
\end{equation}

Observations have determined the masses of the post-SN system, $M_p$ and
$M_c$, the semi-major axis of the present orbit, $A$, and its eccentricity $e$
(see Table~\ref{P1913}). We follow back in time the evolution of the orbit as
determined by gravitational wave emission and derive the semi-major axis,
$A_f$, and eccentricity, $e_f$, just after the second SN explosion.  Finally,
\cite{Kra98}(1998) and \cite{Tay99}(1999) have inferred the inclination of the
pulsar spin with respect to the orbital angular momentum, $\Theta$.  Based on
our understanding of the recycling of the pulsar, the spin axis of the pulsar
(first NS) and the pre-SN orbital angular momentum axis are expected to have
been aligned.  Therefore, this observationally inferred misalignment angle,
$\Theta$, is in fact the tilt angle, $\theta$, of the post-SN orbital plane
relative to that before the explosion.

We use all this information to constrain the (circular) progenitor
system (semi-major axis $A_i$) and the birth kick velocity ${\bf V}_K$
(both magnitude and direction) imparted to the NS during the
SN explosion of the immediate progenitor with mass $M_i$.
In order to do so, we combine the values of $A_f$, $e_f$, and $\theta$
with the results of Section \ref{Motion}, where we discussed the
present position and (transverse) motion of the binary system within our
Galaxy to calculate the center-of-mass velocity, $V_{{\rm CM}}$,
imparted to the system during the explosion as a function of the unknown
radial velocity.  Using Eq.~(\ref{vsvs}) this leads to a curve in the
$A_i$-$M_i$ parameter space, which is constrained by
\begin{equation}\label{cond2}
   \frac{1}{1+e_f} < \alpha < \frac{1}{1-e_f}
\end{equation}
and Eq.~\ref{cond1}. Based on numerical simulations of rapid accretion onto
NSs (\cite{FBH96}1996), \cite{FK97}(1997) have shown that in the pre-SN
progenitor of PSR B1913+16 the orbit must be wide enough so that the pulsar
does not enter a the envelope of the helium star, otherwise it would have
collapsed into a black hole. Therefore, there is an additional constraint
imposed on ($A_i$,$M_i$) determined by the maximum radial extent reached by
the helium stars during their lifetime. Based on stellar evolution
calculations (\cite{Hab85}1985; \cite{WLW95}1995), the maximum radius of
helium stars can be approximated by (\cite{KW96}1996):
\begin{equation}\label{rHe}
\log R_{{\rm He,max}} = \left\{
\begin{array}{ll}
   3.0965-2.013 \log(M_{{\rm He}}/\msun)\;, & M_{{\rm He}}\le 2.5\msun
   \\
   0.0557\;\left[\log(M_{{\rm He}}/\msun)-0.172\right]^{-2.5}\;,
                                            & M_{{\rm He}}> 2.5\msun \;.
\end{array}
\right.
\end{equation}
Since the maximum radius of the immediate progenitor star has to lie within
the Roche lobe of the progenitor system, a lower limit is set on $A_i$ for a
given $M_i$ (see figure 4 in \cite{FK97}1997), which we take into account.


\section{Results and Discussion}

For every allowed pair $(A_i,M_i)$ we can now calculate the kick velocity
${\bf V}_K$ required to form the post-SN system, using Eqs.~\ref{vxvyvz}. We
are interested in its magnitude and its direction with respect to the pre-SN
orbit. We denote the polar angle between the pre-SN orbital angular momentum
and the kick velocity with $\psi_\theta$, and the azimuthal angle with
$\psi_\phi$. (The direction $\psi_\theta=\psi_\phi=0$ points away from the
companion's position at the moment of the SN explosion.)  Hence, for a given
$V_r$ we get an upper and a lower limit on $A_i$, $M_i$, $V_K$, $\psi_\theta$,
and $\psi_\phi$.  A solution for these pre-SN parameters cannot be found for
all values of $V_r$, since values outside the allowed range lead to pairs
$(A_i,M_i)$ that fail to satisfy conditions (\ref{cond1}), (\ref{cond2}), and
(\ref{rHe}).  In this way, we are able to constrain the unknown magnitude of
$V_r$ in addition to the five pre-SN binary and kick characteristics.

The most uncertain parameter in our calculations is the distance to the binary
system and thus its present location in the Galaxy. To account for this
uncertainty we performed our calculations for various distances ranging from
5.9 to 9.7 kpc and plotted the most conservative constraints
(Figs.~\ref{figA1} to \ref{figB2}). Figures \ref{figA1} and \ref{figA2}
correspond to case A, where the pulsar is young ($\sim3$ Myr), with
$\theta=22^\circ$ (prograde) and $\theta=158^\circ$ (retrograde),
respectively. Figures \ref{figB1} and \ref{figB2} give results for case B,
again for the prograde and retrograde cases. Both the mass, $M_i$, of the
exploding helium star and the pre-SN orbital separation, $A_i$, are well
constrained in the retrograde cases A and B and the prograde case B. The kick
magnitude, $V_K$, is better constrained in case B, while both retrograde cases
require higher kick magnitudes, as expected. For all four cases the direction
of the kick is narrowly constrained to be very close to the pre-SN orbital
plane and opposite to the orbital motion of the exploding helium star. In the
retrograde cases the angles $\psi_\phi$ and $\psi_\theta$ are restricted to
extremely narrow ranges.

According to \cite{Kra98}1998 the observational data for the pulse profile
evolution are equally well described by a prograde and retrograde system. He
however argued that a retrograde system would require typically larger and
therefore less probable kick velocities than for the prograde solution, and
therefore considered this case as less likely.  This tendency is indeed
visible in the results of our calculations (Figs.~\ref{figA1} to \ref{figB2}).
More importantly, a comparison between the solid angle covered by the
direction $(\psi_\theta,\psi_\phi)$ in the prograde and retrograde cases (for
each allowed value of $V_r$) indicates that the retrograde cases require a
much higher degree of fine-tuning in the kick direction. For this reason we
consider the retrograde cases to be less probable.

It is important to point out that in all four possible cases the direction of
the kick imparted to the newly-born NS is strongly restricted to be almost in
the orbital plane of the progenitor system ($\psi_\theta \approx 75^\circ -
85^\circ$). Since the spin of the helium star is expected to be aligned with
the pre-SN orbital angular momentum due to the mass transfer in the binary,
our results imply that the kick was given in a direction almost perpendicular
to the spin axis of the NS progenitor.  Such a constraint has interesting and
restrictive implications for the physical mechanism responsible for the kick
in terms of the relevant direction and time scale of the process. Whatever the
details of the mechanism, it must be such that it can allow the kick to be
directed almost perpendicular to the rotation axis of the collapsing star.
Further the time scale relevant to the kick mechanism must be short compared
to the birth spin period of the NS. Otherwise kick components perpendicular to
the NS rotation would be cancelled resulting in a kick parallel to the NS spin
axis, which contradicts the constraints derived here. Our findings are in
agreement with \cite{BT97}1997, \cite{CC98}1998, \cite{DRR99}1999 who argue
that there is no correlation between the pulsar spin and velocity vectors. We
also note that if the kick mechanism is related to the escaping neutrinos
during the collapse, then the spin period of the proto-NS must be longer than
the neutrino-diffusion timescale (a few seconds, \cite{BH96}1996).

The radii of helium stars in late stages of their evolution are not well
defined quantitatively, although the qualitative result of rapid expansion of
relatively low-mass helium stars is common to all the published calculations
(\cite{Hab85}1985; \cite{WLW95}1995). We have explored the sensitivity of our
results by artificially decreasing $R_{{\rm He,max}}$ by a factor of
2. Quantities describing the kick magnitude ($v_K$) and the kick direction
($\psi_\theta$, $\psi_\phi$) turned out to be rather insensitive to changes of
$R_{{\rm He,max}}$. However, the lower bounds on the orbital separation,
$A_i$, and the mass of the Helium star, $M_i$, just before the second SN
explosion proved to change significantly if one reduces $R_{{\rm He,max}}$ by
a factor of 2. The lower bound on $A_i$ decreases by $\sim$50\% for the
$\theta=22^\circ$ cases (Figs.\ \ref{figA1},\ref{figB1}) and $\sim$20\% for
the $\theta=158^\circ$ cases (Figs.\ \ref{figA2},\ref{figB2}). The lower bound
on $M_i$ decreases by $\sim$20\% for the $\theta=22^\circ$ cases, while for
the $\theta=158^\circ$ cases the changes in the lower bound on $M_i$ are less
than 15\%.

In section \ref{Motion} we calculated the kinematic history of the PSR
B1913+16 system based on the model for the Galactic gravitational field of
\cite{KG89}(1989).  In particular, calculations done for the case B
($\tau_{{\rm kin}}\ga35$ Myr) could be sensitive to deviations of the model
from the true Galactic potential. Therefore we repeated all our calculations
using the Galactic potential model published by \cite{Pac90}(1990). The
resulting changes in kick and orbital parameters proved to be below the 10\%
level.  We further tested our assumption of the PSR B1913+16 system being born
at $Z=0$ pc, using in addition $Z=\pm100$ pc as the Galactic height at
birth. Our analysis turned out to be adequately robust against these changes.

Finally, we can use the upper limit on $|V_r|$ to constrain the uncertainty in
the NS-mass measurements.  As pointed out by \cite{DD86}(1986), timing
observations of binary pulsars do not allow the determination of the intrinsic
masses of the binary system. Instead the (intrinsic) masses divided by the
Doppler factor, which to first order is given by $D\simeq 1+V_r/c$, are
measured. Thus our limit of $|V_r|<1200$ km~s$^{-1}$ implies $|M^{{\rm
obs}}/M-1|<0.004$.  This limit on the mass uncertainty could be improved if an
upper bound on the helium-star mass leading to NS formation is obtained
theoretically (from SN collapse calculations) in the future (see
Fig.~\ref{figA1}).

Although some of the limits imposed on the parameters of the immediate
progenitor system and the birth kick are still rather wide, improved values
for the distance to PSR B1913+16 would certainly tighten the constraints.  In
the future an improved distance estimate may be derived from refined
measurements of the relativistic orbital-period decrease (c.f.\
\cite{BB96}1996).  Also, a better estimate for the age of the binary system or
a reliable mass limit on the helium progenitor star from theoretical
considerations combined with the calculations here, would lead to clearly
stricter limits.


\acknowledgements

We greatly appreciate the warm hospitality of the Aspen Center of Physics. We
would also like to thank C. Fryer, D. Psaltis, V. Radhakrishnan, F. Rasio,
J. Taylor, and S. Thorsett for useful discussions.  NW and MK acknowledge the
support by the Max-Planck-Society through the Otto-Hahn-Prize. VK acknowledges
support by the Smithsonian Astrophysical Observatory through a
Harvard-Smithsonian Center for Astrophysics Postdoctoral Fellowship.


\begin{appendix}

\section{The Galactic potential}

It is convenient to use standard cylindrical coordinates $(R,Z,\phi)$ with the
origin at the Galactic center, $R$ is the Galactocentric radius, $\phi$ is the
azimuthal angle and $Z$ is the height above the Galactic plane. The potential
considered here is axi-symmetric about the $Z$-axis and symmetric with respect
to the Galactic plane, i.e.\ $\Phi=\Phi(R,|Z|)$. Following \cite{KI87}(1987)
and \cite{KG89}(1989) the Galactic potential is a sum of three components:
\begin{equation}
   \Phi_{{\rm Gal}} = \Phi_1 + \Phi_2 + \Phi_3 \;,
\end{equation}
where
\begin{equation}
   \Phi_A(R,Z) = -\frac{GM_A}{ \sqrt{\left(a_A+\sum_{i=1}^3\beta_i^A
   \sqrt{Z^2+h_i^A}\right)^2+b_A+R^2}} \;, \qquad A=1,2,3 \;.
\end{equation}
The numerical values for the parameters $M_A$, $a$, $b$, $\beta_i^A$, and
$h_i^A$ are given in Table \ref{GP}. The equations of motion for a star in
the Galactic potential $\Phi=\Phi_{{\rm Gal}}(R,Z)$ are
\begin{equation} \label{eqm}
   \ddot X = \frac{X}{R}\;\frac{\partial\Phi}{\partial R} \;, \qquad 
   \ddot Y = \frac{Y}{R}\;\frac{\partial\Phi}{\partial R} \;, \qquad 
   \ddot Z = \frac{\partial\Phi}{\partial Z}
\end{equation}
In the equations of motion we changed to Cartesian coordinates in order to
avoid the singularity of cylindrical coordinates at $R=0$. The equations of
motion (\ref{eqm}) are then solved numerically using a fourth-order
Runge-Kutta routine (\cite{Pre&al86}1986).

The speed for a circular orbit in the Galactic plane at a distance $R_*$ from
the Galactic center is
\begin{equation}
   V_{{\rm rot}}(R_*) = \sqrt{R_*\frac{\partial\Phi}{\partial R}(R_*,0)} \;.
\end{equation}
At the solar distance, $R_*=8.5$ kpc, the model here gives a circular speed of
221 km/s for the local standard of rest.

\end{appendix}


\newpage 

\figcaption{\label{figage}
Kinematic age, $\tau_{{\rm kin}}$, of the PSR B1913+16 system as a function of
the (unknown) radial velocity, $V_r$, for the three cases A, B, and C.  The
characteristic age of the pulsar, $\tau_{{\rm c}}$, is indicated by a dashed
line. The dotted line indicates the age, $\tau_{{\rm su}}$, inferred from the
spin-up limit during the accretion phase (taken from \cite{ACW98}1998).
}


\figcaption{\label{figz}
Time $\Delta\tau$ required for the PSR B1913+16 system in case B to move from
$z=-300$ pc to $z=+300$ pc devided by its kinematic age as a function of
radial velocity $V_r$. The time spent close to the Galactic plane is less than
8\% of the kinematic age of the PSR B1913+16 system. 
}


\figcaption{\label{figvsys}
Velocity of the PSR B1913+16 system with respect to its Galactic standard of
rest at the moment of birth ($Z=0$ pc) as a function of the (unknown) radial
velocity, $V_r$, for case A (dashed-dotted line) and case B (dashed line).}


\figcaption{\label{figA1} 
Limits on the immediate progenitor (helium-star) mass, $M_i$, the pre-SN
orbital separation, $A_i$, the magnitude, $V_K$, of the kick velocity imparted
to the pulsar companion during the SN explosion, and its direction given by
the two angles, $\psi_\phi$ and $\psi_\theta$, as a function of the (unknown)
radial velocity, $V_r$, for case A and $\theta=22^\circ$. The largest regions
(most conservative results) were obtained for the maximum distance $d$ of 9.7
kpc.}


\figcaption{\label{figA2}
Same as in Fig.~\ref{figA1} for case A and $\theta=158^\circ$.  The largest
regions (most conservative results) were obtained for the maximum distance $d$
of 9.7 kpc.  }


\figcaption{\label{figB1}
Same as in Fig.~\ref{figA1} for case B and $\theta=22^\circ$.  The largest
regions (most conservative results) were obtained for the maximum distance $d$
of 9.7 kpc.  }


\figcaption{\label{figB2}
Same as in Fig.~\ref{figA1} for case B and $\theta=158^\circ$.  The largest
regions (most conservative results) were obtained for the minimum distance $d$
of 6.9 kpc.  }

\newpage 

\begin{table}[h]
\caption{Parameters for the PSR B1913+16 system\label{P1913}}
\begin{tabular}{lll} \hline\hline
R. A. (J2000)           & $\alpha$     & $19^{{\rm h}}15^{{\rm m}}28\fs00$ \\
Decl. (J2000)           & $\delta$     & $16\arcdeg06\arcmin27\farcs4$ \\
Proper motion in R. A. (mas yr$^{-1}$) & $\mu_\alpha=\dot\alpha\cos\delta$
& $-3.27\pm0.35$ \\
Proper motion in Decl. (mas yr$^{-1}$) & $\mu_\delta=\dot\delta$
& $+1.04\pm0.42$ \\
Distance (kpc)          & $d$          & $8.3\pm1.4$            \\[2mm]
Spin period (ms)        & $P$          & 59.03                  \\
Period derivative       & $\dot P$     & $8.63\times10^{-18}$   \\
Characteristic age (Myr)& $\tau_c=P/2\dot P$ & 109              \\[2mm]
Mass of the pulsar ($\msun$)        & $M_p$     & 1.44          \\
Mass of the companion ($\msun$)     & $M_c$     & 1.39          \\
Present semi-major axis ($R_\odot$) & $A$       & 2.80          \\
Present eccentricity                & $e$       & 0.617         \\
Orbital inclination (deg)           & $i$       & 47.2          \\ \hline
\end{tabular}
\end{table}

\newpage 

\begin{table}[h]
\caption{Parameters defining the Galactic potential\label{GP}}
\begin{tabular}{cccccccccc} \hline\hline
      & $M_A$ & $a$ & $b$ & $\beta^A_1$ & $h_1^A$ & $\beta^A_2$ & $h_2^A$ &
      $\beta^A_3$ & $h_3^A$ \\ 
      & ($10^9\msun$) & (kpc) & (kpc) & & (kpc) & & (kpc) & & (kpc) \\ \hline
$A=1$ & 145 & 2.4 & 5.5  & 0.4 & 0.325 & 0.5 & 0.090 & 0.1 & 0.125  \\ 
$A=2$ & 9.3 &  0  & 0.25 &  0  & 0     & 0   & 0     & 0   & 0      \\ 
$A=3$ & 10  &  0  & 1.5  &  0  & 0     & 0   & 0     & 0   & 0      \\ \hline 
\end{tabular}
\end{table}

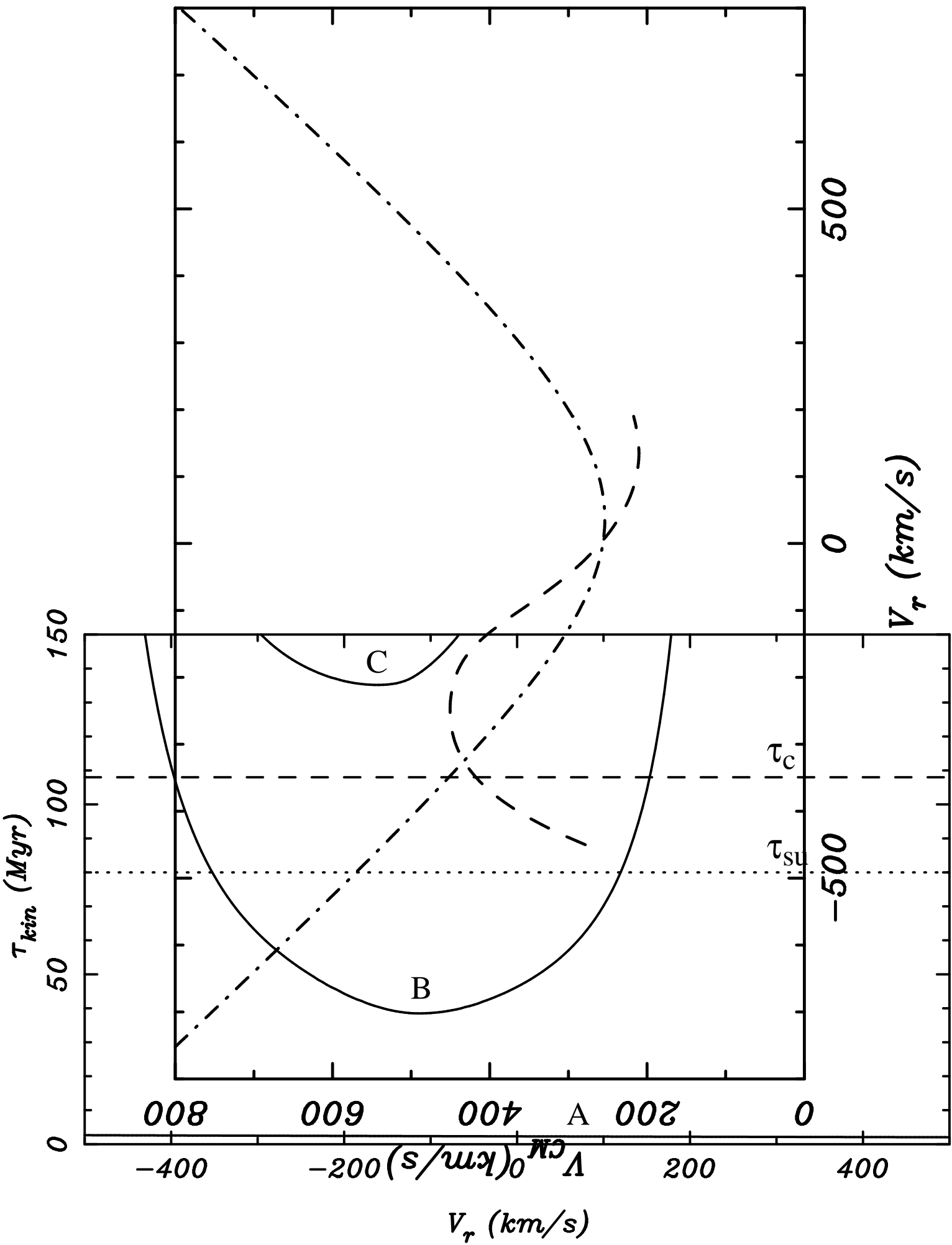
\end{document}